\documentclass[
  reprint,
  superscriptaddress,
  amsmath,
  amssymb,
  aps,
  longbibliography
]{revtex4-2}

\usepackage[utf8]{inputenc}
\usepackage[lining,semibold]{libertine}
\usepackage{amsthm}
\usepackage[libertine,cmintegrals,bigdelims,vvarbb]{newtxmath}

\usepackage{amsmath}
\usepackage{amsfonts}
\usepackage{bm}
\usepackage{mathrsfs}

\usepackage{graphicx}
\usepackage{dcolumn}
\usepackage{booktabs}
\usepackage{siunitx}

\usepackage{xcolor}
\definecolor{webgreen}{rgb}{0,.5,0}
\definecolor{webbrown}{rgb}{.6,0,0}
\definecolor{RoyalBlue}{rgb}{0.0,0.14,0.4}

\usepackage{hyperref}
\hypersetup{
  colorlinks=true,
  linktocpage=true,
  breaklinks=true,
  pdfstartpage=1,
  pdfstartview=FitV,
  pdfpagemode=UseOutlines,
  plainpages=false,
  bookmarksnumbered=true,
  bookmarksopen=true,
  bookmarksopenlevel=1,
  hypertexnames=true,
  pdfhighlight=/O,
  urlcolor=webbrown,
  linkcolor=RoyalBlue,
  citecolor=webgreen,
  pdfauthor={Nahuel Freitas, Timur Aslyamov, Massimiliano Esposito},
  pdfcreator={pdfLaTeX},
  pdfproducer={LaTeX REVTeX}
}

\usepackage[capitalize]{cleveref}

\crefname{appendix}{Appendix}{Appendices}
\Crefname{appendix}{Appendix}{Appendices}

\makeatletter
\def\maketag@@@#1{\hbox{\m@th\normalfont\normalsize#1}}
\makeatother

\begin{document}
\author{Nahuel Freitas}
\email{nfreitas@df.uba.ar}
\affiliation{Departamento de Física, FCEyN, UBA, Pabellón 1, Ciudad Universitaria, 1428 Buenos Aires, Argentina}

\author{Timur Aslyamov}
\email{timur.aslyamov@uni.lu}
\affiliation{Complex Systems and Statistical Mechanics, Department of Physics and Materials Science, University of Luxembourg, 30 Avenue des Hauts-Fourneaux, L-4362 Esch-sur-Alzette, Luxembourg}

\author{Massimiliano Esposito}
\email{massimiliano.esposito@uni.lu}
\affiliation{Complex Systems and Statistical Mechanics, Department of Physics and Materials Science, University of Luxembourg, 30 Avenue des Hauts-Fourneaux, L-4362 Esch-sur-Alzette, Luxembourg}

\title{Emergent Second Law for Time-Dependent Nonequilibrium States}

\date{\today}

\begin{abstract}
For nonautonomous open systems described by macroscopic stochastic thermodynamics, we derive an emergent second law that constrains time-dependent macroscopic fluctuations by the entropy production along the most probable evolution under the time-reversed driving protocol.
We show that this bound can be understood as a macroscopic consequence of the fluctuation theorem: the time reverse of this evolution provides a possible fluctuation path under the forward dynamics.
In the linear-response and slow-driving regime, the bound becomes an equality to first order and generalizes the McLennan--Zubarev formula to the time-dependent probability density of nonautonomous systems with weak nonconservative affinities.
We illustrate our results in a bistable system under sudden quenches and periodic driving.
\end{abstract}
\maketitle

\emph{Introduction---}A central result of equilibrium statistical physics is that the probability of finding a system in a microstate $\bm n$ is determined by its dimensionless potential $\Phi(\bm n)$:
$P_{\mathrm{eq}}(\bm n)=e^{-\Phi(\bm n)}/Z$, where
$Z=\sum_{\bm n}e^{-\Phi(\bm n)}$ is the partition function.
Far from equilibrium, however, the probability $P(\bm n,t)$ depends not only on the system's thermodynamic potential but also on its kinetics and on the driving protocol.
Consequently, the Gibbs form no longer applies, and a general closed analytical expression for $P(\bm n,t)$ is unavailable.

Establishing such expressions and identifying their thermodynamic content is one of the central challenges of nonequilibrium statistical physics.
A partial thermodynamic connection survives close to equilibrium.
For steady states and to linear order in the thermodynamic forces, the McLennan--Zubarev formalism expresses the nonequilibrium correction to the Gibbs probability in terms of the entropy production (EP) accumulated along relaxation trajectories of the reference equilibrium dynamics~\cite{mclennanStatisticalMechanicsSteady1959,zubarev1973nonequilibrium}.
It thus connects the nonequilibrium probability to dissipation, rather than directly to energetics.
This result was later given a rigorous interpretation~\cite{maes2010rigorous,colangeliMeaningfulExpansionDetailed2011} and extended to second order in the thermodynamic forces~\cite{komatsuExpressionStationaryDistribution2008}.
Beyond linear response, however, time-symmetric dynamical activity also enters~\cite{colangeliMeaningfulExpansionDetailed2011}, showing that EP alone does not generally determine the probability.
For small state spaces, stationary probabilities can be computed using linear algebra~\cite{aslyamov2024nonequilibrium, frezzato2024steady} and graph-theoretic methods~\cite{hill1966studies,king1956schematic,schnakenberg1976network}.
Beyond stationary states, exact analytical expressions for $P(\bm n,t)$ are known in few cases such as one-step processes~\cite{karlin1957birthdeath}, special classes of reaction networks~\cite{lubensky2010equilibriumlike,anderson2010product} 
and a few integrable interacting particle systems such as ASEP and TASEP~\cite{schutz1997exact}.
For generic systems far from equilibrium, however, neither a universal closed-form expression for $P(\bm n,t)$ nor a general thermodynamic characterization of this probability in terms of energy or EP is known.

Further progress is possible in the thermodynamic limit, in which a large parameter $V\to\infty$ while the rescaled state $\bm c\equiv\bm n/V$ remains intensive.
The finite-system probability $P(\bm n,t)$ then induces the macroscopic probability density
$p(\bm c,t)\equiv V^d P(V\bm c,t)$, where $d$ is the dimension of the macroscopic state space.
This limit describes, among others, chemical reaction networks (CRNs)~\cite{andersonContinuousTimeMarkov2011,lazarescuLargeDeviationsDynamical2019,tamano2026universal,qian2021stochastic}, nonlinear electronic circuits~\cite{freitasStochasticThermodynamicsNonlinear2021,Gao2021Aug,freitasReliabilityEntropyProduction2022,gopalLargeDeviationsTheory2022}, and interacting spin systems~\cite{herpichCollectivePowerMinimal2018,HerpichGian2020}.

To accommodate time-dependent (TD) driving from the outset, we allow the potential $\Phi(\bm n,t)$ to depend parametrically on an external control protocol.
At each time $t$, freezing the control parameters at their instantaneous values defines an equilibrium reference density.
In the thermodynamic limit, this density takes the large-deviation form~\cite{touchette2009large}
\begin{align}
\label{eq:p-eq}
    p_{\mathrm{eq}}(\bm c,t)
    \asymp
    e^{-V[\phi(\bm c,t)-\phi_{\mathrm{eq}}(t)]}\,,
\end{align}
where
$\phi(\bm c,t)\equiv\lim_{V\to\infty}V^{-1}\Phi(V\bm c,t)$
and $\bm x^\text{eq}_t$ is a minimum of $\phi(\bm c,t)$, with
$\phi_{\mathrm{eq}}(t)\equiv
\phi(\bm x^\text{eq}_t,t)$.
Here $a_V\asymp b_V$ denotes logarithmic equivalence, $\lim_{V\to\infty}V^{-1} \ln (a_V/b_V)=0$.
The system follows this family of equilibrium densities in the quasistatic-driving limit.
At finite driving speed, it provides a convenient time-dependent equilibrium reference for the actual nonequilibrium density.

More generally, the nonequilibrium density also obeys a large-deviation form,
\begin{align}
\label{eq:P-LD}
    p(\bm c,t)\asymp e^{-V I(\bm c,t)}\,,
\end{align}
where $I(\bm c,t)\geq0$ is the time-dependent rate function.
Within a selected attraction basin, its minimum defines the most probable macroscopic state $\bm x(t)$:
\begin{align}
\label{eq:rate-fun-most-prob}
    I(\bm x(t),t)=0\,,
    \ \ \ \
    \nabla I(\bm x(t),t)=0\,,
    \ \ \ \
    \nabla^2 I(\bm x(t),t) \succeq 0\,.
\end{align}
We reserve $\bm x(t)$ for this most probable state and denote by
$\bm x_t$ a solution of the nonautonomous deterministic dynamics
\begin{equation}
\label{eq:det_dyn}
    d_t\bm x_t=\bm F(\bm x_t,t)\,,
\end{equation}
corresponding to an arbitrary initial condition $\bm x_0$. 
The two coincide, $\bm x(t)=\bm x_t$, when 
\begin{equation}
\label{eq:det_dyn_IC}
    \bm x_0 = \bm x(0) \equiv 
    \operatorname*{arg\,min}_{\bm c\in B_\alpha(0)}
    I(\bm c,0)\,.
\end{equation}
Here $\bm F(\bm c,t)$ is the thermodynamically consistent drift derived in \cref{eq:drift}, and $B_\alpha(0)$ is the selected attraction basin.
In a multistable system, this construction is understood branch-wise: the choice of $B_\alpha(0)$ selects the corresponding deterministic trajectory.

For autonomous dynamics, $\bm F=\bm F(\bm c)$, the stationary rate function $I_\text{ss}(\bm c)$ is related to the macroscopic EP rate $\dot\sigma(\bm c)$ through the \textit{emergent second law} (ESL)~\cite{freitasEmergentSecondLaw2022}:
\begin{equation}
    \int_0^{t_f}dt\,
    \dot\sigma(\bm x_t) \equiv \sigma[\bm x_{t_f}] \geq
    I_\text{ss}(\bm x_{0})
    - I_\text{ss}(\bm x_{t_f})\,.
    \label{eq:intro_esl}
\end{equation}
Moreover, the stationary rate function decreases along the deterministic evolution~\cite{gangLyapounovFunctionStationary1986},
\begin{align}
\label{eq:d_tI-ss}
    d_t I_\text{ss}(\bm x_t) \leq 0\,.
\end{align}
Consequently, \cref{eq:intro_esl} implies the second law of thermodynamics, $\sigma[\bm x_{t_f}] \geq 0$.
The ESL was originally derived from the adiabatic--nonadiabatic decomposition of EP in the thermodynamic limit~\cite{freitasEmergentSecondLaw2022}.
It also holds for Langevin dynamics~\cite{Santolin2025Feb}, while related dissipation bounds were obtained earlier for reaction--diffusion systems~\cite{gaveauDissipationEnergyInformation1998}.

In this Letter, we ask whether the connection between macroscopic fluctuations and dissipation extends to time-dependent nonequilibrium states.
Our first main result is a generalization of the ESL to nonautonomous systems, which bounds the time-dependent rate function in terms of its initial value and the entropy produced along the most probable evolution under the time-reversed driving protocol.
This bound has a direct path-space interpretation as a macroscopic consequence of the fluctuation theorem. Time-reversing the most probable evolution under the reversed protocol yields a possible fluctuation path under the forward protocol, with a dynamical cost set by the entropy produced along the reversed evolution.
Our second main result shows that, in the linear-response and slow-driving regime, the bound becomes an equality to first order and generalizes the McLennan--Zubarev formula to the time-dependent probability density of nonautonomous systems with weak nonconservative affinities.

\emph{Setup---}We consider stochastic systems described microscopically by continuous-time Markov jump processes.
Let $\bm n\in\mathbb Z^d$ denote a microscopic state and let $P(\bm n,t)$ be the probability of finding the system in that state at time $t$.
Its evolution is governed by the master equation
\begin{align}
\partial_t P(\bm n,t)
=
\sum_\rho\Big[
&R_\rho(\bm n-\bm\Delta_\rho,t)
P(\bm n-\bm\Delta_\rho,t)
-
R_\rho(\bm n,t)P(\bm n,t)
\Big]\,,
\label{eq:master_eq}
\end{align}
where $R_\rho(\bm n,t)$ is the rate of the transition
$\bm n\to\bm n+\bm\Delta_\rho$.
For every transition $\rho$, we assume that the reverse transition $-\rho$ is also allowed, with
$\bm\Delta_{-\rho}=-\bm\Delta_\rho$.

Measuring entropy in units of $k_\mathrm{B}$, local detailed balance identifies the reservoir entropy change associated with $\rho$ as
\begin{align}
\Sigma_\rho(\bm n,t)
&\equiv
\ln
\frac{
R_\rho(\bm n,t)
}{
R_{-\rho}(\bm n+\bm\Delta_\rho,t)
}
\nonumber\\
&=
-\Phi(\bm n+\bm\Delta_\rho,t)
+\Phi(\bm n,t)
+a_\rho(\bm n,t)\,,
\label{eq:ldb}
\end{align}
where $\Phi(\bm n,t)$ is the dimensionless state potential and
$a_\rho(\bm n,t)$ is a nonconservative (NC) affinity satisfying $a_{-\rho}(\bm n+\bm\Delta_\rho,t) =-a_\rho(\bm n,t)$.
At fixed $t$, we call the dynamics detailed balance when $\forall \rho: a_\rho(\bm n,t)=0$.
The mean EP rate is
\begin{align}
\label{eq:micro-EPR}
\langle\dot\Sigma\rangle
\equiv
\sum_{\rho>0,\bm n}
J_\rho(\bm n,t)
\left[
\Sigma_\rho(\bm n,t)
+
\ln
\frac{
P(\bm n,t)
}{
P(\bm n+\bm\Delta_\rho,t)
}
\right]
\geq0\,,
\end{align}
where
$J_\rho(\bm n,t) \equiv R_\rho(\bm n,t)P(\bm n,t)-R_{-\rho}(\bm n+\bm\Delta_\rho,t)
P(\bm n+\bm\Delta_\rho,t)$
is the probability current, and $\rho>0$ denotes one orientation of each pair of reverse transitions.

We now take the macroscopic limit \cite{falasco2025macroscopic}.
We assume that the increments $\bm\Delta_\rho$ remain finite as $V\to\infty$, while the state potential and transition rates are extensive:
$ r_\rho(\bm c,t) \equiv \lim_{V\to\infty} R_\rho(V\bm c,t)/V$,
$ \phi(\bm c,t) \equiv \lim_{V\to\infty} \Phi(V\bm c,t)/V$.
The NC affinities remain intensive,
$ a_\rho(\bm c,t) \equiv \lim_{V\to\infty}a_\rho(V\bm c,t)$,
and the macroscopic probability density is
$p(\bm c,t)=V^dP(V\bm c,t)$.
At leading order in $V$, \cref{eq:master_eq} becomes
\begin{align}
\label{eq:macro-master}
\partial_t p(\bm c,t)
=
V\sum_\rho\Big[
&r_\rho\left(\bm c-\frac{\bm\Delta_\rho}{V},t\right)
p\left(\bm c-\frac{\bm\Delta_\rho}{V},t\right)
\nonumber\\
&-
r_\rho(\bm c,t)p(\bm c,t)
\Big]\,.
\end{align}
Substituting the large-deviation form \cref{eq:P-LD} and expanding to leading order in $V^{-1}$ gives the Hamilton--Jacobi equation
\begin{equation}
\partial_t I(\bm c,t)
=
\sum_\rho r_\rho(\bm c,t)
\left[
1-e^{\bm\Delta_\rho\cdot\nabla I(\bm c,t)}
\right].
\label{eq:HJ}
\end{equation}

Multiplying \cref{eq:macro-master} by $\bm c$ and summing over the microscopic states gives
$ d_t\langle\bm c\rangle
= \sum_\rho \left\langle r_\rho(\bm c,t)\right\rangle \bm\Delta_\rho$,
where
$\langle O(\bm c)\rangle
\equiv
\int d\bm c\,p(\bm c,t)O(\bm c)$ and
$\int d\bm c
\equiv
\lim_{V\to\infty}
V^{-d}\sum_{\bm n}$.
Within a selected branch, the probability density concentrates on the most probable state $\bm x(t)$ as $V\to\infty$, so that
$\lim_{V\to\infty}
\langle O(\bm c)\rangle
= O(\bm x(t))$.
The mean dynamics therefore reduces to the deterministic dynamics in \cref{eq:det_dyn}, with drift
\begin{equation}
\bm F(\bm c,t)
\equiv
\sum_\rho r_\rho(\bm c,t)\bm\Delta_\rho\,.
\label{eq:drift}
\end{equation}

Combining \cref{eq:HJ,eq:drift} with $e^u\geq1+u$, we obtain
\begin{align}\label{eq:dtI}
\partial_t I(\bm c,t)
&+
\bm F(\bm c,t)\cdot\nabla I(\bm c,t) \\
&=
\sum_\rho r_\rho(\bm c,t)
\left[
1+\bm\Delta_\rho\cdot\nabla I(\bm c,t)
-
e^{\bm\Delta_\rho\cdot\nabla I(\bm c,t)}
\right] \leq0\,. \nonumber
\end{align}
Consequently, along any deterministic trajectory $\bm x_t$,
\begin{align}
d_t I(\bm x_t,t)\leq 0\,.
\label{GenLyapunov}
\end{align}
This establishes a new nonautonomous Lyapunov property: the time-dependent rate function is nonincreasing along every deterministic trajectory.
When $\bm x_0=\bm x(0)$ is the most probable initial state selected in \cref{eq:det_dyn_IC}, the trajectory remains at the instantaneous minimum of the rate function and 
\begin{equation}
d_t I(\bm x(t),t)=0\,.
\end{equation}
\Cref{GenLyapunov} generalizes the stationary property in \cref{eq:d_tI-ss}.

Using the macroscopic limit of \cref{eq:ldb},
\begin{align}
\label{eq:macro_ldb}
\sigma_\rho(\bm c,t) &\equiv \lim_{V\to \infty}\Sigma_\rho(V\bm c,t) \\
&= \ln \frac{r_\rho(\bm{c}, t)}{r_{-\rho}(\bm{c}, t)}
\!=
- \bm{\Delta}_\rho \cdot \nabla\phi(\bm c,t) + a_\rho(\bm c,t)\nonumber\,,
\end{align}
and the probability density concentration along the deterministic trajectory, we find that the macroscopic limit of the scaled mean EPR in \cref{eq:micro-EPR} reads $\lim_{V\to\infty} \langle \dot \Sigma\rangle/V=\dot \sigma (\bm{x}(t),t)$, where
\begin{align}
\label{eq:macro-EPR}
    \dot \sigma (\bm{c},t) \equiv \sum_\rho r_\rho(\bm{c},t) \sigma_\rho(\bm{c},t) \geq 0\,.
\end{align}
The function $\dot\sigma(\bm c,t)$ can be evaluated along any deterministic trajectory $\bm x_t$ and decomposed as
\begin{align}
\label{eq:epr-decomp-forward}
    \dot{\sigma}(\bm{x}_t, t)
    &=  \dot\sigma_\text{nc}(\bm{x}_t, t) - \bm{F}(\bm{x}_t, t) \cdot \nabla \phi(\bm{x}_t, t) \nonumber \\
    &=  \dot\sigma_\text{nc}(\bm{x}_t, t) + \partial_t\phi(\bm{x}_t, t) - d_t\phi(\bm{x}_t, t) \,.
\end{align}
Here, $\dot{\sigma}_\text{nc}(\bm c,t)\equiv\sum_\rho r_\rho(\bm c,t)a_\rho(\bm c,t)$ and $\partial_t\phi(\bm c,t)$ are, respectively, the NC and TD contributions to the EPR. The former can be expressed in terms of products of fundamental thermodynamic forces and fluxes and the latter, in isothermal systems, to the Jarzynski work \cite{rao2018conservation}.

\textit{First main result: time-dependent ESL---}We express the rate function as
\begin{align}
\label{eq:I-split}
    I(\bm{c},t) & = \phi(\bm{c}, t) + g(\bm{c}, t) - [\phi(\bm x(t), t) + g(\bm x(t), t)]\,,
\end{align}
where $g(\bm{c},t)$ is the nonequilibrium correction to the instantaneous potential~\cite{freitas2021linear}.
At equilibrium, $g(\bm c,t)=0$ and, within the selected branch, $\bm x(t)=\bm x^\text{eq}_t$, where $\bm x^\text{eq}_t$ is an instantaneous minimum of the potential satisfying
$\nabla\phi(\bm x^\text{eq}_t,t)=0$.
Thus, $\bm x^\text{eq}_t$ is an instantaneous fixed point of the frozen deterministic dynamics when $a_\rho(\bm c,t)=0$.
It becomes the actual evolving state only in the quasistatic-driving limit.
In general, $g(\bm{c},t)$ and $a_\rho(\bm{c},t)$ are related through
\begin{equation}
    \partial_t I(\bm{c}, t) = \sum_{\rho} r_\rho(\bm{c}, t) \left( 1 - 
    e^{-a_\rho(\bm{c},t)}
    e^{-\bm{\Delta}_\rho \cdot \nabla g(\bm{c},t)}\right)\,,
    \label{eq:I_t-W-g}
\end{equation}
whose derivation from \cref{eq:HJ} is provided in \cref{sec:I_t-W-g}.

Using $e^{x} \geq 1+x$ and \cref{eq:drift} in \cref{eq:I_t-W-g}, we obtain
\begin{equation}
    \partial_t I(\bm{c}, t) - \bm{F}(\bm{c},t)\cdot \nabla g(\bm{c},t) \leq \dot \sigma_\text{nc}(\bm{c}, t)\,.
    \label{eq:prefinal_step}
\end{equation}
Subtracting $\bm{F}(\bm{c},t)\cdot\nabla\phi(\bm{c},t)$ from both sides and using \cref{eq:I-split,eq:epr-decomp-forward}, we find
\begin{equation}
    \partial_t I(\bm{c},t) - \bm{F}(\bm{c},t) \cdot \nabla I(\bm{c}, t) \leq \dot \sigma(\bm{c},t)\,.
    \label{eq:gen_esl}
\end{equation}
At the most probable state, $\bm c=\bm x(t)$, \cref{eq:rate-fun-most-prob} gives $\nabla I(\bm x(t),t)=0$, while \cref{eq:HJ} gives $\partial_t I(\bm x(t),t)=0$.
Consequently, \cref{eq:gen_esl} reduces to the second law, $\dot\sigma(\bm x(t),t)\geq0$.

It is more interesting to evaluate \cref{eq:gen_esl} along an
arbitrary backward deterministic trajectory $\hat{\bm{x}}_t$ defined by
\begin{equation}
d_t\hat{\bm{x}}_t
=
-\bm F\bigl(\hat{\bm{x}}_t,t\bigr),
\qquad
0\leq t\leq t_f.
\label{eq:rev_dyn}
\end{equation}
When traversed backward from $t_f$ to $0$, this trajectory is the
most probable macroscopic evolution under the time-reversed driving
protocol, starting from the arbitrary initial point
$\hat{\bm{x}}_{t_f}$.
Evaluating \cref{eq:gen_esl} along $\hat{\bm{x}}_t$ gives
\begin{equation}
    d_t I(\hat{\bm{x}}_t, t) \leq \dot \sigma(\hat{\bm{x}}_t, t)\,,
    \label{eq:gen_esl_mat}
\end{equation}
where
$
d_t I(\hat{\bm{x}}_t, t) = \partial_t I(\hat{\bm{x}}_t, t) - \bm{F}(\hat{\bm{x}}_t, t) \cdot \nabla I(\hat{\bm{x}}_t, t)
$
is the total derivative along the $t$-parametrized backward trajectory.
Using \cref{eq:epr-decomp-forward}, we also have
\begin{align}
\label{eq:epr-decomp-backward}
    \dot{\sigma}(\hat{\bm{x}}_t,t) &=  \dot\sigma_\text{nc}(\hat{\bm{x}}_t, t) - \bm{F}(\hat{\bm{x}}_t, t) \cdot \nabla \phi(\hat{\bm{x}}_t, t) \nonumber\\
    &=  \dot\sigma_\text{nc}(\hat{\bm{x}}_t,t) - \partial_t \phi(\hat{\bm{x}}_t,t) + d_t\phi(\hat{\bm{x}}_t,t) \,.
\end{align}
Integrating \cref{eq:gen_esl_mat}, we derive the \textit{time-dependent} ESL:
\begin{equation}
    I(\hat{\bm{x}}_{t_f}, t_f) - I(\hat{\bm{x}}_0, 0) \leq \sigma[\hat{\bm x}_0] \,,
    \label{eq:esl-nonauto}
\end{equation}
with 
\begin{align}
\label{eq:entropy-prod-reversed}
    \sigma[\hat{\bm x}_0] 
    = \sigma_\text{nc}[\hat{\bm x}_0] - \sigma_d[\hat{\bm x}_0] + \phi(\hat{\bm x}_{t_f},t_f) - \phi(\hat{\bm x}_0,0)\,,
\end{align} 
where
\begin{align}
&\sigma[\hat{\bm x}_0] \equiv  \int_{0}^{t_f}dt \dot \sigma(\hat{\bm{x}}_t, t) \label{IntDefofEP}\,, \\
\sigma_\text{nc}[\hat{\bm x}_0] \equiv \int_{0}^{t_f} & \dot \sigma_\text{nc}(\hat{\bm{x}}_t, t) dt \;,\quad  
\sigma_d[\hat{\bm x}_0] \equiv \int_{0}^{t_f}\partial_t \phi (\hat{\bm{x}}_t, t) dt \nonumber \;.
\end{align} 

\Cref{eq:esl-nonauto} also has a direct path-space interpretation.
When traversed from $t_f$ to $0$, $\hat{\bm x}_t$ is the typical
macroscopic evolution under the reversed protocol, starting from
$\hat{\bm x}_{t_f}$.
The same curve traversed from $0$ to $t_f$ is a fluctuation path
under the forward protocol, starting from $\hat{\bm x}_0$.
The conditional fluctuation theorem compares the probability of a
narrow tube around this forward path, conditioned on its initial
point $\hat{\bm x}_0$, with that of the reversed tube, conditioned on
its initial point $\hat{\bm x}_{t_f}$.
Since the reversed trajectory is typical, its conditional probability
carries no large-deviation cost.
Consequently, at leading exponential order, the conditional
probability of the forward path scales as
$\exp\{-V\sigma[\hat{\bm x}_0]\}$.
Multiplying by the probability
$\exp\{-V I(\hat{\bm x}_0,0)\}$ of its initial point shows that this
forward history has the total large-deviation exponent
$I(\hat{\bm x}_0,0)+\sigma[\hat{\bm x}_0]$.
Since the final density sums over all forward histories ending at
$\hat{\bm x}_{t_f}$, its rate function cannot exceed this exponent,
yielding \cref{eq:esl-nonauto}.
The inequality is saturated when this history is the dominant one.
A complete derivation is given in \cref{sec:FT-ESL}.

\textit{Special cases of time-dependent ESL---}For
\textit{autonomous dynamics}, the reversed protocol coincides with
the forward protocol.
Consequently,
$\bm x_0=\hat{\bm x}_{t_f}$ and $\bm x_{t_f}=\hat{\bm x}_0$,
where $\bm{x}_t$ is the solution of the forward deterministic
dynamics initialized at $\bm x_0=\hat{\bm x}_{t_f}$.
Equivalently, $\hat{\bm{x}}_t$, when traversed from $0$ to
$t_f$, is the time reverse of the deterministic relaxation
$\bm{x}_t$.
\Cref{eq:esl-nonauto} then simplifies to
\begin{equation}
    I(\bm x_0, t_f) - I(\bm{x}_{t_f}, 0) \leq \sigma_\text{nc}[\bm x_{t_f}] + \phi(\bm x_0) - \phi(\bm x_{t_f})\,.
    \label{eq:td_esl-auto}
\end{equation}
\Cref{eq:td_esl-auto} is a time-dependent generalization of the ESL
from \cite{freitasEmergentSecondLaw2022}.
In steady state, the rate function is time independent but is
evaluated at two different points along a deterministic trajectory.
Therefore, the left-hand side of \cref{eq:td_esl-auto} becomes
$I_\text{ss}(\bm x_0)-I_\text{ss}(\bm x_{t_f})$, recovering
\cref{eq:intro_esl}.

For \textit{periodically driven} systems, Floquet theory applied to
the master equation in \cref{eq:master_eq} shows that, whenever the
dynamics is asymptotically stable, $p(\bm c,t)$ and hence
$I(\bm c,t)$ become $T$-periodic at long times~\cite{barato2018periodic}.
The corresponding most probable trajectory $\bm x(t)$ is also
$T$-periodic, whereas the arbitrary backward trajectory entering the
ESL generally is not.
For $t_0$ in the periodic regime, let $\hat{\bm x}_t$ start from the
arbitrary initial point $\hat{\bm x}_{t_0+T}$ under reversed driving
and satisfy
\begin{equation}
d_t\hat{\bm{x}}_t
=
-\bm F(\hat{\bm{x}}_t,t),
\qquad
t_0\leq t\leq t_0+T,
\end{equation}
Using $I(\hat{\bm x}_{t_0+T},t_0+T)
=I(\hat{\bm x}_{t_0+T},t_0)$ in \cref{eq:esl-nonauto} gives
\begin{equation}
I(\hat{\bm{x}}_{t_0+T},t_0)
-
I(\hat{\bm{x}}_{t_0},t_0)
\leq
\int_{t_0}^{t_0+T}dt\,
\dot{\sigma}(\hat{\bm{x}}_t,t).
\label{eq:bound_periodic}
\end{equation}

\begin{figure*}
    \centering
    \includegraphics[width=\linewidth]{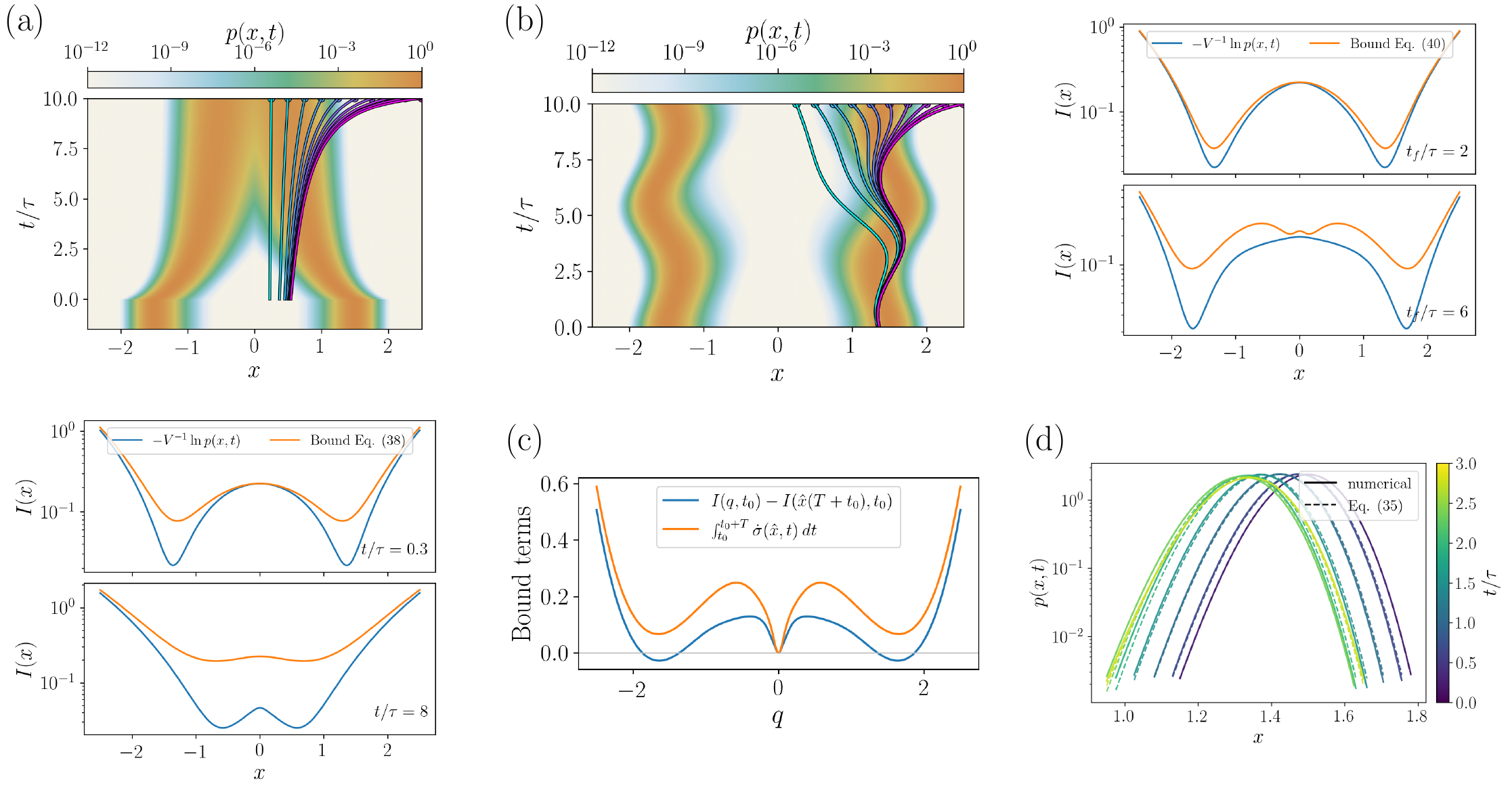}
    \caption{
Relaxation after an instantaneous quench of the double-well potential (a) and a periodically driven double-well system (b--d).
(a) Top: Evolution of the probability density $p(c,t)\simeq VP(n,t)$ after the control parameter is switched at $t=0$ from $\lambda^-=1.5$ to $\lambda^+=0$, with $\alpha=0.04$ and $V=200$.
The density is obtained by numerical integration of the master equation in \cref{eq:master_eq}, truncated to $\lvert n\rvert\leq n_c=2.5V$.
Colored curves show backward deterministic trajectories $\hat{x}_t$.
Bottom: Scaled self-information $I(c,t)=-V^{-1}\ln p(c,t)$ at $t/\tau=0.3$ and $t/\tau=8$, compared with the autonomous ESL bound in \cref{eq:td_esl-auto} for $\sigma_\text{nc}=0$.
(b) Left: Evolution of the probability density $p(c,t)\simeq VP(n,t)$ under the protocol $\lambda(t)=a-b\sin(2\pi t/T)$, with $a=1.5$, $b=0.3$, $T=2\pi\tau$, $\alpha=0.04$, and $V=200$.
The system is initialized in the equilibrium state corresponding to $\lambda(0)=1.5$.
Colored curves show backward deterministic trajectories $\hat{x}_t$.
Right: Scaled self-information $I(c,t)=-V^{-1}\ln p(c,t)$ at $t_f/\tau=2$ and $t_f/\tau=6$, compared with the time-dependent ESL bound in \cref{eq:bound-rate-periodic}.
(c) Periodic-state test of the Floquet ESL bound in \cref{eq:bound_periodic}: the rate-function difference $I(\hat{x}_{t_0+T},t_0)-I(\hat{x}_{t_0},t_0)$ is bounded by the EP accumulated along the backward deterministic trajectory over one driving period.
(d) Comparison between the numerical probability density and the nonautonomous McLennan--Zubarev formula in \cref{eq:McLennan} at several times during the first three relaxation times.
}
\label{fig:panel}
\end{figure*}
\textit{Second main result---}We now consider the regime in which
the NC affinities and the driving speed are both small.
We quantify the expansion by the dimensionless parameter
\begin{equation}
\varepsilon
\sim
\max_\rho\sup_{\bm c,t}|a_\rho(\bm c,t)|
+
\tau_{\mathrm{rel}}
\sup_{\bm c,t}\left|
\partial_t\left[\phi(\bm c,t)-\phi_{\mathrm{eq}}(t)\right]
\right|
\ll 1,
\label{eq:small-parameter}
\end{equation}
where $\tau_{\mathrm{rel}}$ is a deterministic relaxation time of
the reference detailed-balance dynamics~\cite{mandal2016slow}.
The suprema are understood over the dynamically relevant basin.
Subtracting the instantaneous minimum makes the slow-driving condition
independent of state-independent shifts of the potential.
We assume a regular perturbative expansion in which
$g(\bm c,t)=\mathcal O(\varepsilon)$ and
$\bm\Delta_\rho\cdot\nabla g(\bm c,t)=\mathcal O(\varepsilon)$
throughout the relevant basin.
We also assume that the rates depend smoothly on the affinities, so that
$r_\rho(\bm c,t)=r_\rho^0(\bm c,t)+\mathcal O(\varepsilon)$, with 
$r_\rho^0(\bm c,t)\equiv r_\rho(\bm c,t)|_{\bm a=0}$.
We therefore linearize \cref{eq:I_t-W-g} by neglecting terms of
order
$\mathcal O[(a_\rho+\bm{\Delta}_\rho\cdot\nabla g)^2]
=\mathcal O(\varepsilon^2)$:
\begin{equation}
\partial_t I^1(\bm c,t)
=
\bm F^0(\bm c,t)\cdot\nabla g(\bm c,t)
+
\dot{\sigma}_{\mathrm{nc}}^1(\bm c,t),
\label{eq:I_t-W-g-linear}
\end{equation}
where 
$I(\bm c,t)=I^1(\bm c,t)+\mathcal O(\varepsilon^2)$,
$ \bm F^0(\bm c,t) \equiv \sum_\rho \bm{\Delta}_\rho r_\rho^0(\bm c,t)$,
and $ \dot{\sigma}_{\mathrm{nc}}^1(\bm c,t) \equiv \sum_\rho r_\rho^0(\bm c,t)a_\rho(\bm c,t)$.
Here and below, the superscript $1$ denotes a quantity consistently
truncated through first order, rather than its first-order correction
alone, while $\bm F^0$ is the instantaneous detailed-balance reference
drift.

At this order, the difference between the two sides of
\cref{eq:prefinal_step} is $\mathcal O(\varepsilon^2)$, so the
time-dependent ESL becomes an identity. Integrating this identity
along the backward detailed-balance dynamics gives, for an equilibrium
initial state and at leading exponential order, the time-dependent probability density
\begin{align}
p(\bm q,t_f) \asymp
p_{\mathrm{eq}}(\bm q,t_f) 
e^{-V \sigma^1[\hat{\bm x}^0_0]},
\label{eq:McLennan}
\end{align}
where the first-order correction in the exponent is
\begin{align}\label{eq:first-order-integrated-contributions}
&\hspace{0.5cm}\sigma^1[\hat{\bm x}^0_0] =
\sigma_{\mathrm{nc}}^1[\hat{\bm x}^0_0]
-
\sigma_d^1[\hat{\bm x}^0_0]
+
\phi_{\mathrm{eq}}(t_f)
-
\phi_{\mathrm{eq}}(0) \\
&\sigma_{\mathrm{nc}}^1[\hat{\bm x}^0_0]
\equiv
\int_0^{t_f}dt\,
\dot{\sigma}_{\mathrm{nc}}^1(\hat{\bm x}^0_t,t)\,,\quad
\sigma_d^1[\hat{\bm x}^0_0]
\equiv
\int_0^{t_f}dt\,
\partial_t\phi(\hat{\bm x}^0_t,t). \nonumber
\end{align}
Here $\hat{\bm x}^0_t$ is the backward reference trajectory satisfying
$d_t\hat{\bm x}^0_t=-\bm F^0(\hat{\bm x}^0_t,t)$ and
$\hat{\bm x}^0_{t_f}=\bm q$.
Thus, for each $\bm q$, applying \cref{eq:McLennan} only requires
integrating the detailed-balance reference dynamics backward from
$\bm q$ and evaluating the two integrals in
\cref{eq:first-order-integrated-contributions}.
The derivation is given in \cref{sec:McLennan}.
In a multistable system, \cref{eq:McLennan} holds branch-wise.
For a selected basin $B_\alpha(t_f)$, it gives the conditional
large-deviation density on that branch.

\Cref{eq:McLennan} generalizes the McLennan--Zubarev formula to
nonautonomous macroscopic systems whose dynamics need not satisfy
detailed balance~\cite{mclennanStatisticalMechanicsSteady1959,
zubarev1973nonequilibrium}; see also
\cite{freitasEmergentSecondLaw2022}.
For autonomous dynamics, $\sigma_d^1=0$ and
$\phi_{\mathrm{eq}}(t_f)=\phi_{\mathrm{eq}}(0)$, so that
$\sigma^1=\sigma_{\mathrm{nc}}^1$ in \cref{eq:McLennan}.
In the steady-state limit, the backward reference trajectory can be
viewed, in the opposite direction, as the forward detailed-balance
relaxation initialized at $\bm q$. This recovers the standard
McLennan formula for Markov jump processes~\cite{maes2010rigorous}.

\emph{Applications---}We apply our results to a simple model of an
equilibrium memory: a one-dimensional system
($n\in\mathbb Z$) with the time-dependent quartic potential
$\Phi(n,t)= \frac{\alpha}{V^3}
\left[n^2-n_0(t)^2\right]^2$
and transitions $n\to n\pm1$ with rates
\begin{equation}
R_\pm(n,t)
=
\frac{V}{\tau}
\exp\left[
-\frac{\Phi(n\pm1,t)-\Phi(n,t)}{2}
\right],
\end{equation}
where $\tau$ is a characteristic time scale.
The dynamics satisfies detailed balance, so the NC affinities vanish.
Departures from equilibrium are generated solely by changes in the
time-dependent protocol $\lambda(t)\equiv n_0(t)/V$ and the ensuing
relaxation.

As a first protocol, we consider an instantaneous quench and study
the subsequent relaxation.
For $t<0$, the system is in equilibrium at $\lambda^-$. Since the
minimum of the potential vanishes, its initial rate function is
$I(c,0^-)=\phi^-(c)$, with
$\phi^-(c)\equiv\alpha[c^2-(\lambda^-)^2]^2$.
At $t=0$, the parameter changes instantaneously from $\lambda^-$ to
$\lambda^+$.
The probability does not change during the quench, and therefore
$I(c,0)=I(c,0^-)=\phi^-(c)$.
The system subsequently undergoes autonomous relaxation towards the
new equilibrium, with post-quench potential
$\phi^+(c)\equiv\alpha[c^2-(\lambda^+)^2]^2$.
For the simulations, we take a finite but large $V$, numerically solve
the truncated master equation for $P(n,t)$, and plot
$p(c,t)\simeq VP(n,t)$ in \cref{fig:panel}(a, top).
In \cref{fig:panel}(a, bottom), we plot the bound
\cref{eq:td_esl-auto} with $\sigma_\text{nc}=0$ as
\begin{align}
I(q,t_f)
\leq
\phi^+(q)
+
\phi^-(x^0_{t_f})
-
\phi^+(x^0_{t_f}),
\end{align}
where $x^0_t$ is the deterministic detailed-balance trajectory satisfying
\begin{equation}
d_t x^0_t
=
-\frac{2}{\tau}
\sinh\left[
\frac{
\left.\partial_c\phi^+(c)\right|_{c=x^0_t}
}{2}
\right],
\qquad
x^0_0=q.
\end{equation}

As a second protocol, we consider periodic TD driving,
$\lambda(t)=a-b\sin(2\pi t/T)$.
The evolution of the probability density and the backward
deterministic trajectories $\hat x_t$, starting from
$\hat x_{t_f}=q$ under reversed driving, are shown in
\cref{fig:panel}(b, left).
In \cref{fig:panel}(b, right), we apply the ESL bound
\cref{eq:esl-nonauto}.
Since the initial state is at equilibrium and the instantaneous
potential minimum vanishes, $I(c,0)=\phi(c,0)$, and the dynamics satisfies detailed balance,
\cref{eq:entropy-prod-reversed} gives
\begin{equation}
I(q,t_f)
\leq
\phi(q,t_f)
-
\int_0^{t_f}dt\,
\frac{\partial\phi(\hat x_t,t)}{\partial\lambda}
\dot{\lambda}(t).
\label{eq:bound-rate-periodic}
\end{equation}

At long times, the system reaches its Floquet state, where we test
the periodic ESL bound \cref{eq:bound_periodic}; see
\cref{fig:panel}(c).
Because the protocol is slow, the first-order nonautonomous
McLennan--Zubarev formula, \cref{eq:McLennan}, is expected to be
accurate up to $\mathcal O(\varepsilon^2)$ in the rate function; see
\cref{fig:panel}(d).

For both protocols, the bound generally becomes looser as the
observation interval increases and more entropy is produced.
Nevertheless, even at long times it reproduces qualitative features
of the actual self-information and therefore remains informative.
The generalized McLennan--Zubarev formula provides both qualitative
and quantitative agreement with the numerical probability density.

\emph{Conclusion---}For a broad class of macroscopic stochastic systems, our results
establish a thermodynamic connection between time-dependent
nonequilibrium probability densities and dissipation.
The ESL constrains the full large-deviation rate function far from
equilibrium and provides a practical route to constructing
approximations to stationary and time-dependent nonequilibrium states
from EP and deterministic trajectories alone, without solving the full
stochastic dynamics.
Near equilibrium and under slow driving, this construction becomes
exact to first order and generalizes the stationary
McLennan--Zubarev formula to nonautonomous dynamics, treating slow
time-dependent driving and weak nonconservative affinities on the
same footing.
In this sense, these results may be viewed as a fluctuation--dissipation relation for the entire large-deviation function.
The path-space derivation further reveals that the fluctuation theorem
and time reversal lie at the core of the ESL: the most probable
evolution under reversed driving, when traversed in the opposite
direction, provides a fluctuation path under the forward dynamics.
Together, these results provide thermodynamic insight into one of the
central problems of nonequilibrium statistical physics, namely the
characterization of nonequilibrium probability distributions.

\begin{acknowledgments}
T.A. and M.E. are funded by the Fonds National de la Recherche (FNR), Luxembourg: project ThermoElectroChem (C23/MS/18060819) and NEQPHASETRANS (C24/MS/18933049), respectively.
\end{acknowledgments}

\begin{center}
  \large \bf End Matter
\end{center}
\vspace{-1cm}
\appendix
\crefalias{section}{appendix}

\section{Derivation of \cref{eq:I_t-W-g}}
\label{sec:I_t-W-g}

Using \cref{eq:I-split}, we first rewrite \cref{eq:HJ} as
\begin{align*}
\partial_t I(\bm c,t)
&=
\sum_\rho r_\rho(\bm c,t)
\left[
1-
e^{\bm{\Delta}_\rho\cdot\nabla\phi(\bm c,t)}
e^{\bm{\Delta}_\rho\cdot\nabla g(\bm c,t)}
\right]
\\
&=
\sum_\rho r_\rho(\bm c,t)
\left[
1-
\frac{r_{-\rho}(\bm c,t)}{r_\rho(\bm c,t)}
e^{a_\rho(\bm c,t)}
e^{\bm{\Delta}_\rho\cdot\nabla g(\bm c,t)}
\right]
\\
&=
\sum_\rho r_\rho(\bm c,t)
-
\sum_\rho r_{-\rho}(\bm c,t)
e^{a_\rho(\bm c,t)}
e^{\bm{\Delta}_\rho\cdot\nabla g(\bm c,t)}
\\
&=
\sum_\rho r_\rho(\bm c,t)
\left[
1-
e^{-a_\rho(\bm c,t)}
e^{-\bm{\Delta}_\rho\cdot\nabla g(\bm c,t)}
\right].
\end{align*}
In the second equality, we used \cref{eq:macro_ldb}; in the last,
we changed the summation index $\rho\to-\rho$ and used
$a_{-\rho}(\bm c,t)=-a_\rho(\bm c,t)$ and
$\bm{\Delta}_{-\rho}=-\bm{\Delta}_\rho$.
This yields \cref{eq:I_t-W-g}.

\section{Path-space derivation of the time-dependent ESL}
\label{sec:FT-ESL}

Here we show that \cref{eq:esl-nonauto} can be understood as a
macroscopic consequence of the fluctuation theorem.
Let $\mathcal{P}_{\mathrm F}[\omega|\bm n_0]$ denote the conditional
probability of a microscopic path $\omega$ under the forward
protocol, and let
$\mathcal{P}_{\mathrm R}[\hat{\omega}|\bm n_f]$ denote the
conditional probability of its time reverse $\hat{\omega}$ under the
reversed protocol.
The transition rates of the reversed process are
\begin{equation}
R_\rho^{\mathrm R}(\bm n,s)
\equiv
R_\rho(\bm n,t_f-s).
\end{equation}
Here $s\in[0,t_f]$ denotes the time parameter of the reversed process,
while $t=t_f-s$ is the corresponding time of the forward protocol.
Thus, if $\omega$ starts from $\bm n_0$ and ends at $\bm n_f$,
then $\hat{\omega}$ starts from $\bm n_f$ and ends at $\bm n_0$.

Suppose that $\hat{\omega}$ undergoes transitions $\rho_j$ at times
$s_j$.
Local detailed balance, \cref{eq:ldb}, gives the conditional
fluctuation theorem
\begin{align}
\ln
\frac{
\mathcal{P}_{\mathrm R}[\hat{\omega}|\bm n_f]
}{
\mathcal{P}_{\mathrm F}[\omega|\bm n_0]
}
&=
\sum_j
\ln
\frac{
R_{\rho_j}(\bm n(s_j^-),t_f-s_j)
}{
R_{-\rho_j}(\bm n(s_j^+),t_f-s_j)
}
\equiv
\Sigma_{\mathrm{res}}^{\mathrm R}[\hat{\omega}] .
\label{eq:conditional-FT}
\end{align}
The waiting-time contributions cancel because the path and the
protocol are reversed simultaneously.
The quantity $\Sigma_{\mathrm{res}}^{\mathrm R}$ is the entropy
change in the reservoirs along the reversed-process path.

Let $\hat{\bm{x}}_t$ be the backward trajectory starting from the
initial point $\hat{\bm{x}}_{t_f}$ under the reversed protocol.
When traversed from $t_f$ to $0$, it is the most probable macroscopic
evolution under that protocol.
When traversed from $0$ to $t_f$, the same curve is a fluctuation
path under the forward protocol.

To account for the reaction currents that generate this curve, let
$\hat{\gamma}$ denote the typical reaction-resolved macroscopic path
under the reversed protocol, consisting of $\hat{\bm x}_t$ and its
typical empirical transition currents, and let
$\gamma=\Theta\hat{\gamma}$ denote its time reverse.
Let
$\mathbb P_{\mathrm R}^{V,\delta}
[\hat{\gamma}|\hat{\bm x}_{t_f}]$
and
$\mathbb P_{\mathrm F}^{V,\delta}
[\gamma|\hat{\bm x}_0]$
denote the conditional probabilities of paired tubes of width
$\delta$ around these reaction-resolved paths.
The macroscopic limit is taken at fixed $\delta$, followed by the
shrinking-tube limit $\delta\to0$.
Since $\hat{\gamma}$ is the typical reaction-resolved path under the
reversed dynamics,
\begin{equation}
\lim_{\delta\to0}\lim_{V\to\infty}
\frac{1}{V}
\ln
\mathbb P_{\mathrm R}^{V,\delta}
[\hat{\gamma}|\hat{\bm x}_{t_f}]
=0.
\end{equation}

For microscopic paths converging to this typical reversed
reaction-resolved path, \cref{eq:macro_ldb,eq:macro-EPR} give
\begin{align}
\lim_{V\to\infty}
\frac{
\Sigma_{\mathrm{res}}^{\mathrm R}[\hat{\omega}]
}{V}
&=
\int_0^{t_f}ds\,
\dot{\sigma}
\bigl(\hat{\bm{x}}_{t_f-s},t_f-s\bigr)
\nonumber\\
&=
\int_0^{t_f}dt\,
\dot{\sigma}
\bigl(\hat{\bm{x}}_t,t\bigr)
=
\sigma[\hat{\bm x}_0].
\label{eq:macro-conditional-FT}
\end{align}
Because the empirical transition currents are also constrained, this
convergence is uniform within the shrinking tube: all microscopic
paths in the tube have the same scaled EP at leading order.

Rewriting \cref{eq:conditional-FT} as
$\mathcal P_{\mathrm F}=\mathcal P_{\mathrm R}
\exp(-\Sigma_{\mathrm{res}}^{\mathrm R})$
and integrating over the paired tubes gives
\begin{align}
\frac{
\mathbb P_{\mathrm F}^{V,\delta}
[\gamma|\hat{\bm x}_0]
}{
\mathbb P_{\mathrm R}^{V,\delta}
[\hat{\gamma}|\hat{\bm x}_{t_f}]
}
&=
\left\langle
\exp\left\{
-\Sigma_{\mathrm{res}}^{\mathrm R}[\hat{\omega}]
\right\}
\right\rangle_{\mathrm R,\delta}
\nonumber\\
&\asymp
\exp\{-V\sigma[\hat{\bm x}_0]\}.
\label{eq:tube-conditional-FT}
\end{align}
Here $\langle\cdot\rangle_{\mathrm R,\delta}$ denotes the normalized
average over reversed microscopic paths in the tube around
$\hat{\gamma}$.
The first equality is exact at finite $V$ and $\delta$, while the
second holds at leading exponential order when $V\to\infty$ followed
by $\delta\to0$.

Defining the conditional forward action of a reaction-resolved
macroscopic path by
\begin{equation}
\mathcal A_{\mathrm F}[\gamma]
\equiv
-\lim_{\delta\to0}\lim_{V\to\infty}\frac{1}{V}
\ln
\mathbb P_{\mathrm F}^{V,\delta}
[\gamma|\bm z(0)],
\end{equation}
we thus have
\begin{equation}
\mathcal A_{\mathrm F}[\gamma]
=
\sigma[\hat{\bm x}_0].
\end{equation}

The contraction principle gives
\begin{align}
I(\hat{\bm x}_{t_f},t_f)
=
\inf_{\substack{
\bm c_0,\,
\gamma'=(\bm z,\bm j):\\
\bm z(0)=\bm c_0,\,
\bm z(t_f)=\hat{\bm x}_{t_f}
}}
\left[
I(\bm c_0,0)
+
\mathcal A_{\mathrm F}[\gamma']
\right],
\label{eq:path-contraction}
\end{align}
where the infimum is over all admissible reaction-resolved
macroscopic paths $\gamma'=(\bm z,\bm j)$ with the specified
endpoints.
Evaluating it on $\gamma=\Theta\hat{\gamma}$ yields
\begin{align}
I(\hat{\bm x}_{t_f},t_f)
&\leq
I(\hat{\bm{x}}_0,0)
+
\sigma[\hat{\bm x}_0],
\end{align}
which is \cref{eq:esl-nonauto}.
The inequality is saturated when $\gamma$ realizes the minimum in
\cref{eq:path-contraction}.

\section{Derivation of \cref{eq:McLennan}}
\label{sec:McLennan}

In the linear-response and slow-driving regime,
we consider the backward detailed-balance reference trajectory
\begin{equation}
d_t\hat{\bm{x}}^0_t
=
-\bm F^0(\hat{\bm{x}}^0_t,t),
\qquad
\hat{\bm{x}}^0_{t_f}=\bm q.
\label{eq:backward-reference-dynamics}
\end{equation}
Using \cref{eq:I-split,eq:I_t-W-g-linear} along this trajectory gives
the first-order identity
\begin{align}
d_t I^1(\hat{\bm{x}}^0_t,t)
&=
\dot{\sigma}_{\mathrm{nc}}^1(\hat{\bm x}^0_t,t)
-\bm F^0(\hat{\bm x}^0_t,t)\cdot
\nabla\phi(\hat{\bm x}^0_t,t)
\nonumber\\
&=
\dot{\sigma}_{\mathrm{nc}}^1(\hat{\bm x}^0_t,t)
-\partial_t\phi(\hat{\bm x}^0_t,t)
+d_t\phi(\hat{\bm x}^0_t,t).
\label{Ident}
\end{align}
For the equilibrium initial condition,
$I^1(\bm c,0)=\phi(\bm c,0)-\phi_{\mathrm{eq}}(0)$.
Integrating \cref{Ident} and using
\cref{eq:first-order-integrated-contributions}, the dependence on the
initial point $\hat{\bm x}^0_0$ cancels between this initial rate
function and the boundary contribution from $d_t\phi$, yielding
\begin{align}
\label{eq:rate-fun-McLennan}
I^1(\bm q,t_f)
&=
\phi(\bm q,t_f)-\phi_{\mathrm{eq}}(0)
+\sigma_{\mathrm{nc}}^1[\hat{\bm x}^0_0]
-\sigma_d^1[\hat{\bm x}^0_0]
\\
&=
I_{\mathrm{eq}}(\bm q,t_f)
+\sigma_{\mathrm{nc}}^1[\hat{\bm x}^0_0]
-\sigma_d^1[\hat{\bm x}^0_0]
+\phi_{\mathrm{eq}}(t_f)-\phi_{\mathrm{eq}}(0)\,,\nonumber
\end{align}
where $I_{\mathrm{eq}}(\bm q,t_f)=
\phi(\bm q,t_f)-\phi_{\mathrm{eq}}(t_f)$ by \cref{eq:p-eq}.
Combining \cref{eq:rate-fun-McLennan,eq:P-LD,eq:p-eq} gives
\cref{eq:McLennan}.

The equilibrium baseline appearing here is fixed thermodynamically:
\begin{equation}
\phi_{\mathrm{eq}}(t)
=
-\lim_{V\to\infty}\frac{1}{V}\ln Z_{\mathrm{eq}}(t),
\qquad
Z_{\mathrm{eq}}(t)
=
\sum_{\bm n}e^{-\Phi(\bm n,t)}.
\label{eq:equilibrium-partition-baseline}
\end{equation}
For a branch-wise construction, the sum is restricted to the selected
basin at the large-deviation level. Hence the initial equilibrium
condition and the equilibrium partition functions fix the additive
constant in \cref{eq:rate-fun-McLennan}.

\bibliography{references}
\end{document}